\documentclass[aps,prb,twocolumn,showpacs,amsmath,amssymb,floatfix]{revtex4}

\usepackage{graphicx}
\usepackage{dcolumn}
\usepackage{bm}
\usepackage{epstopdf}

\begin{document}
\bibliographystyle{prsty}

\title{ RKKY Interaction in Graphene from Lattice Green's Function}
\author{M. Sherafati and S. Satpathy}
\affiliation{Department of Physics, University of Missouri, Columbia, Missouri 65211, USA
}
\date{\today}

\begin{abstract}

We study the exchange interaction $J$ between two magnetic impurities in graphene (the RKKY interaction) by directly computing the lattice Green's function for the tight-binding band structure for the honeycomb lattice. The method allows us to compute $J$ numerically for much larger distances than can be handled by finite-lattice calculations as well as for small distances.
In addition, we rederive the analytical long-distance  behavior of $J$ for linearly dispersive bands  and find corrections to the oscillatory factor that were previously missed in the literature.
The main features of the RKKY interaction in graphene are that
unlike the $J \propto (2k_FR)^{-2} \sin (2k_FR) $ behavior of an ordinary 2D metal in the long-distance limit, $J$ in graphene falls off as $1/R^3$, shows the $1 + \cos ((K-K').R)$-type oscillations with additional phase factors depending on the direction, and exhibits a ferromagnetic interaction for moments on the same sublattice and an antiferromagnetic interaction for moments on the opposite sublattices as required by particle-hole symmetry.
The computed $J$ with the full band structure agrees with our analytical results in the long-distance limit including the oscillatory factors with the additional phases.

 \end{abstract}

\maketitle

\section{Introduction}
Graphene has attracted considerable attention recently due to its linear energy dispersion, where the excitations are massless Dirac Fermions, which could lead to physical behavior different from that of the standard two-dimensional systems. The Ruderman-Kittel-Kasuya-Yosida (RKKY) interaction is the exchange interaction between two magnetic impurities mediated by the conduction electrons of the host and is a fundamental quantity of interest.\cite{RKKY,RKKY-2D}
Earlier works on the RKKY interaction in graphene\cite{Saremi, Brey, Vozmediano, Dugaev}  have mostly used a continuum model with a linearly dispersive band structure with two Dirac cones at the corners of the Brillouin zone or have used exact diagonalization on a finite size lattice.\cite{Annica}
As pointed out by Saremi,\cite{Saremi, Bunder} for a bipartite lattice with nearest-neighbor interactions, particle-hole symmetry leads to a ferromagnetic interaction on the same sublattice and
an antiferromagnetic interaction for the opposite sublattices.

The continuum limit, linearly-dispersive Dirac cone approximation is expected to be good in the long-distance limit $R \rightarrow \infty$ and allows an analytical solution. However, it requires the use of a cutoff function, without which the contributions from the higher energy states add up to produce a diverging and oscillatory result. If a sharp energy cutoff is used, it produces a $J$ that oscillates with distance violating the particle-hole symmetry.
To address this problem, a cutoff function approach\cite{Saremi} has been used, where the higher-momentum contributions are damped out slowly, with the length scale of damping taken to infinity as the limiting case. While this approach yields a reasonable result, it is not {\it a priori} obvious if some systematic error is not introduced by such a procedure.
In fact,  using exact diagonalization on finite lattices, Black-Schaffer\cite{Annica} extracted $J$ values which differed from Saremi's results both in the prefactors and, for $J_{AB}$, in the oscillatory factor as well. On the other hand, the finite lattice calculations in turn suffer from the deficiency that the distance between the  impurities can not be too large and extra interactions between the moments  get introduced due to the supercell geometry.

In order to address these issues, we use a direct computation of the lattice Green's functions to compute the RKKY interaction with the full tight-binding band structure. The method allows us to numerically calculate the RKKY interaction for very large distances, which is impractical to obtain from the finite lattice calculations.  In addition, we obtain  analytical results for the long-distance behavior of $J$ by using an approach slightly different from that of Saremi,\cite{Saremi} which allows us to obtain the proper phase factors in the RKKY oscillations that were missed in the previous works. We note that for obtaining the proper oscillatory factors of $J$, it is important to include carefully the phase factors of the electronic wave functions around the Dirac points.

\section{Formulation}

We consider the tight-binding Hamiltonian for graphene including a contact interaction with the magnetic centers, viz.,
\begin{equation}
{\cal H}  =
 \sum_{\langle ij \rangle \sigma}t_{ij} c^{\dagger}_{i\sigma}c_{j\sigma} + h. c.
- \lambda \sum_p \vec S_p \cdot \vec s_p,
\label{hamil}
\end{equation}
 where $i$ is the combined site-sublattice index, $\langle ij \rangle$ denotes summation over distinct pairs of nearest-neighbor sites, $\sigma$ denotes the electron spin, the $p$ summation runs over the magnetic centers, and
 $\vec s_p = (\hbar/2) \sum_{\mu\nu} c^\dag_{p \mu} \vec \tau_{\mu\nu} c_{p \nu}$ is the itinerant electron spin density. The results can be easily generalized if the hopping integrals $t_{ij}$ are retained beyond the nearest neighbor. However, direct numerical computations showed that neglecting the higher neighbor terms does not change $J$ significantly, since the magnitudes of the hopping beyond the nearest-neighbors are relatively small in graphene.\cite{Nanda-Graphene} The nearest-neighbor hopping parameter in graphene is
$ t = - 2.56$ eV as obtained by fitting the tight-binding bands to the density-functional band calculations.\cite{Nanda-Graphene}

\begin{figure}
\centering
\includegraphics[width=6.0cm]{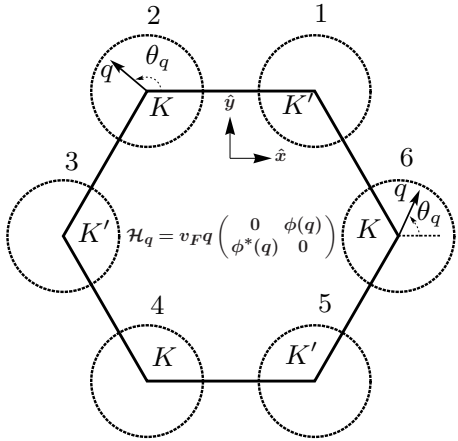}
\caption{The Hamiltonian near the Dirac points in the first Brillouin zone with  $\phi(q) = - e^{i(\pi/3-\theta_q)}, \  e^{i(\pi/3 + \theta_q)}, \ e^{- i \theta_q}, \ - e^{i(2 \pi/3+\theta_q)}, \ e^{i(2 \pi/3 - \theta_q)}, \ -e^{ i \theta_q} $, for the first through the sixth Dirac points ($K_1$ to $K_6$)  as labelled in the figure. The small momentum $q$ is the deviation from the corresponding Dirac point ($k=K_D+q$) and  $\theta_q $ is the polar angle of $q$ with respect to $K_1 - K_2$ chosen as the $\hat x$ direction as shown in the figure. These phase factors are important as they determine  the oscillatory behavior of $J_{AB}$ as discussed in the text.
}
\label{Fig-H}
\end{figure}

  In the basis of the Bloch functions ($c^{\dagger}_{k\alpha\sigma}= N^{-1/2} \sum_i  e^ {ik.(R_i+\tau_\alpha)}     c^{\dagger}_{i \alpha \sigma}  $ )
  of the two sublattices, $\alpha = A$ or $B$, the unperturbed Hamiltonian is given by
\begin{equation}
{\cal H}_k=
 \left( \begin{array}{cc}
0 &  f(k)\\
f^*(k) & 0
\end{array}\right),
\end{equation}
where  $f(k) = t \ (e^{i  k \cdot  d_1} + e^{i  k \cdot  d_2} + e^{i  k \cdot  d_3} ) $.
  The Hamiltonian expression near a Dirac point takes the form $f(k) = v_F q \ \phi (q)$,  where $\phi (q)$ is a phase, different at different Dirac points as indicated in Fig. \ref{Fig-H}, and $q$ is the deviation of the Bloch momentum $k$ from the neighboring Dirac point, $q = k - K_D$. Diagonalizing the Hamiltonian, one finds a linear dispersion near the Dirac points, viz., $E_q =  \pm |v_F| q$ with the Fermi velocity $v_F \equiv 3 a t /2$, which is defined to be negative throughout this paper, $t$ being negative, and here $a$ is the  bond length.

\begin{figure}
\centering
\includegraphics[width=5.0cm]{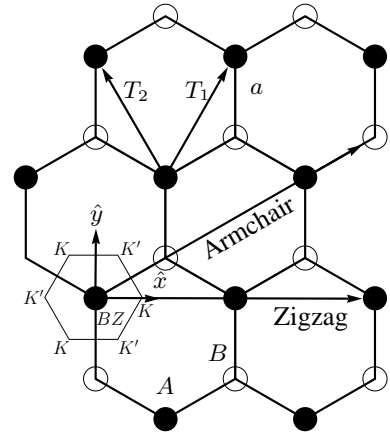}
\caption{The graphene honeycomb lattice with two different sublattices, shown as full and open circles. The figure shows the orientation of the Brillouin zone and two common directions in the direct lattice (zigzag and armchair).
}
\label{H}
\end{figure}

In the linear response theory, the exchange interaction may be obtained by first computing the perturbed wave functions due to the  magnetic impurity located at the origin from  the Lippmann-Schwinger equation $| \Psi \rangle = | \Psi^0 \rangle + G V | \Psi \rangle$, from which the energy due to the second magnetic impurity located at $R$ is computed from the first-order perturbation theory, so that  $E(R) = \langle \Psi | H_{int} (R) | \Psi \rangle$.
For  a contact interaction between the magnetic impurity and the conduction electrons,  ${\cal H}_{int} = -\lambda \ \vec S \cdot \vec s \ \delta (\vec r)$, $E(R)$ may be written in the Heisenberg form
 \begin{equation}
 E(R) = J \vec S_1 \cdot \vec S_2,
 \label{Heisenberg}
 \end{equation}
 where the exchange integral  $J = (\lambda^2 \hbar^2/4) \ \chi (0,R)$. The susceptibility
 $\chi (r,r^\prime) =\delta n(r) / \delta V(r')$
 is written in terms of the unperturbed retarded Green's functions
\begin{equation}
 \chi (r,r^\prime) = - \frac{2}{\pi} \int^{E_F} dE\  { \rm Im} [G^0 (r,r^\prime, E)
 G^0 (r^\prime, r, E)].
\end{equation}
Here $G^0$ is the Green's function for a single spin channel and is spin independent, while in the definition of the susceptibility, $\delta V(r')$ is a spin-independent perturbation and $\delta n(r)$ is the change in the total charge density {\it including} both spins.

This result can be easily extended to the case of graphene to yield
 \begin{equation}
    \chi_{\alpha \beta} (r,r^\prime) =
- \frac{2}{\pi} \int^{E_F} dE \
 {\rm Im} [G^0_{\alpha \beta} (r,r^\prime, E)
G^0_{\beta \alpha  } (r^\prime, r, E)],
\label{chi}
\end{equation}
where $\alpha, \beta$ are the sublattice indices, A or B,  $r, r^\prime$ denote the lattice positions of the two magnetic centers located on the sublattice $\alpha$ and $\beta$, respectively, and the sublattice susceptibility is as usual $ \chi_{\alpha \beta} (r,r^\prime) \equiv  \delta n_\alpha(r) / \delta V_\beta(r')$.
The expression for the susceptibility is obtained by noting that the charge density is given by
$n_\alpha(r) = - \frac{2}{\pi} \int^{E_F} dE\   {\rm Im} \ G_{\alpha\alpha} (r,r, E)
 $ and obtaining the charge difference $ \delta n_\alpha(r)$ induced by the perturbation $\delta V_\beta(r')$
 from the Dyson equation $G=G^0 + G^0VG$.
The exchange interaction $J_{\alpha \beta} (R)$  between impurities located at the sites ($\alpha, 0$) and ($\beta, R$) is then given by
\begin{equation}
J_{\alpha \beta} (R) = \frac{\lambda^2 \hbar^2}{4}  \chi_{\alpha \beta} (0,R).
\end{equation}
The calculation of the exchange interaction   thus boils down to the computation of the lattice Green's functions and a quadrature over the energy following Eq. \ref{chi}.

\section {Calculation of the Green's Functions}

We compute the real-space Green's function by two different approaches, viz., the direct integration method and the recursive technique of Horiguchi.\cite{Horiguchi} In the first method, the real-space Green's functions are calculated by numerically integrating the momentum-space Green's functions
\begin{equation}
G^0_{\alpha \beta} (r,r^\prime, E) = \frac{1}{\Omega_{BZ}}\int d^2k e^{i k \cdot (r-r^\prime)}   G^0_{\alpha \beta} (k, E),
\label{GF}
\end{equation}
where
\begin{eqnarray}
G^0 (k,E)
& =  &
 \left( \begin{array}{cc}
G^0_{AA}  &  G^0_{AB}\\
G^0_{BA} & G^0_{BB}
\end{array}\right)
\equiv  (E+i \eta - {\cal H}_k)^{-1}
  \nonumber \\
&=& \frac{E+i \eta + {\cal H}_k}{(E+i \eta )^2 - f^*(k) f(k)}.
\label{GF-matrix}
\end{eqnarray}

The Brillouin zone integral in Eq. \ref{GF} was evaluated by taking up to $N_k = 10^6$ $k$-points in the full Brillouin zone and a small value for the infinitesimal parameter $\eta =0.005$ is used. This method is straightforward and computationally robust but slow, while the Horiguchi recursive technique is fast, but it has stability problems\cite{Berciu} for larger distances.

It is worth noting the symmetry properties of the Green's functions, which immediately follow from the above equations and the expression for $ {\cal H} _k$, viz., that  $G_{AA} (R,0,E) = G_{BB} (R,0,E)$ and $G_{AB} (0,R,E) = G_{BA} (R,0,E)$, leading to the results, which is also obvious on physical grounds:
\begin{equation}
J_{AA} (R) = J_{BB} (R)\ {\rm and} \  J_{AB} (R) = J_{BA} (-R).
\end{equation}

In the second method, the Horiguchi recursive technique,\cite{Horiguchi}
 the Green's functions for the honeycomb lattice is expressed in terms of those for the triangular lattice, which in turn are expressed in terms of the elliptic integrals. For example, the expression for the on-site Green's function is given by
 \begin{equation}
 G_{AA}^0 (0,0,z) = \frac{z}{4 \pi} g (z' ) \tilde{K} (k(z' )),
 \label{Eq-G00}
 \end{equation}
where $z = E + i \eta$, $z' = (z^2-3) / 2$,
$g(z') = 8 ((2z'+3)^{1/2}-1)^{-3/2}    ( (2z'+3)^{1/2}+3)^{-1/2}  $,
$k(z') = 2^{-1} g(z') (2z'+3)^{1/4}$,
$\tilde {K} (k) = K (k)$ for $ {\rm Im}\  k < 0$
and $K (k) + 2 i K (\sqrt {1-k^2})$ for $ {\rm Im}\  k > 0$, and $K(k)$ is the elliptic integral
 \begin{equation}
 K(k) = \int_0^{\pi/2} \frac{1}{(1-k^2 \sin ^2 \theta)^{1/2}} \ d\theta.
 \end{equation}
 The elliptic integral with complex modulus was evaluated following established procedures and the Arithmetic-Geometric Mean Method\cite{Ref-Elliptic1,Ref-Elliptic2} and $\eta = 10^{-6}$ was used.

\begin{figure}
\centering
\includegraphics[width=8.0cm]{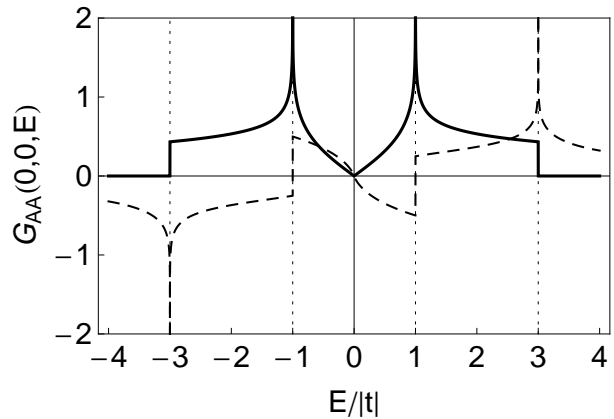}
\caption{On-site Green's function Re $G_{AA}^0(0,0,E)$ (dashed line) and $- {\rm Im }   \ G_{AA}^0(0,0,E)$ (full line)  computed from Eq. \ref{Eq-G00}.
}
\label{G00}
\end{figure}

The computed result for the on-site Green's function obtained from Eq. \ref{Eq-G00} is shown in Fig. \ref{G00}, and one sees the familiar density-of-states, which is proportional to the imaginary part.  Similarly, one can compute the Green's functions for a few lattice vectors $R$ (specifically, $R \equiv (l,m) = (0,0), (2,0), {\rm and} \ (4,0)$, for the triangular lattice), from which the remaining Green's functions for both the triangular as well as the honeycomb lattice can be computed using the recursion relations.

\begin{figure}
\centering
\includegraphics[width=8.0cm]{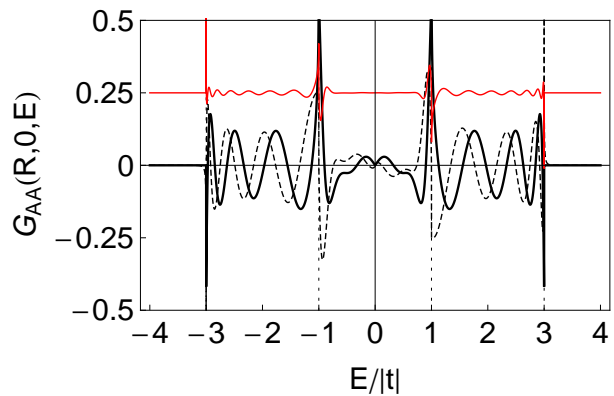}
\caption{(Color online) Real (dashed line) and Imaginary (full line) parts of the Green's function $G_{AA}^0(R,0,E)$ with $R = 7 \sqrt {3} a (1,0)$ obtained from the Horiguchi recursive method. Red line shows the product of the Real and Imaginary parts, which is displaced by +0.25 along the y-axis.
}
\label{G180}
\end{figure}

In Fig. \ref{G180}, we have plotted the Green's function $G_{AA}^0(R,0,E)$ with a specific $R = 7 \sqrt {3} a (1,0)$ along the zigzag direction and also the product of the real and the imaginary parts, which enters as the integrand in the calculation of $J$ (Eq. \ref{chi}), to be integrated over the occupied states between $E = -3$ and zero. As seen from the figure, the integrand is a rapidly oscillating function, with a small net result for $J$.

\section{Exchange Interaction}

Before we present the results of the full calculations, we derive the long-distance behavior of $J$ with the linearly dispersive band structure. This is reasonable as the long-distance behavior is necessarily determined by the small momentum states, for which the linear-band approximation is excellent. As has been pointed out in the literature,\cite{Saremi,Annica} the RKKY interaction shows oscillations as a function of distance because of the interference between charge densities originating from the two Dirac points in the Brillouin zone. However, there is no consensus regarding the form of the oscillation and we find important differences from the earlier results.

\subsection{Magnetic impurities on the same sublattice}
To derive the long-distance oscillatory behavior of the RKKY interaction, we first obtain the Green's functions for small energies (which necessarily determine the long distance behavior) and use these to evaluate the integral in the expression Eq. \ref{chi} for the susceptibility. For small energies, the contribution to the integral in Eq. \ref{GF} comes from the two Dirac points in the BZ, so that the equation becomes
\begin{eqnarray}
&&G^0_{\alpha \beta} (R,0, E) = \frac{1}{\Omega_{BZ}}\int d^2q \ e^{i q \cdot R}  \nonumber \\
&\times &[  e^{i K \cdot R} G^0_{\alpha \beta}  (q+K, E) +  e^{i K^\prime \cdot R} G^0_{\alpha \beta}  (q+K^\prime, E)],
\label{G-RE}
\end{eqnarray}
where $K$ and $K^\prime$ are the two Dirac points and $R$ is the distance between the two magnetic centers.

For $\alpha = \beta$, i.e., if the two sublattices are the same,  these two Green's functions in the kernel become the same, viz.,
$G^0_{AA} (q, E)
= (E+i \eta ) ((E+i \eta )^2 - v_F^2 q^2)^{-1}$
as seen from Eq. \ref{GF-matrix}. The exponential factor may be expanded (Jacobi-Anger expansion\cite{GRHandbook}) in terms of the Bessel's functions
\begin{equation}
 e^ {i q \cdot R} = J_0 (qR) + 2 \sum_{n=1}^\infty i^n J_n (qR) \cos [n (\theta_q -\theta_R)]
\label{Jacobi}
\end {equation}
 and then integrating over $\theta_q$, we get the result
\begin{equation}
G^0_{AA} (R,0, E) = (e^{i K \cdot R}  +  e^{i K^\prime \cdot R}) g_{AA} (R,E),
\end{equation}
where
\begin{equation}
g_{AA} (R,E) = \frac {2 \pi}{\Omega_{BZ}} \int_0^{q_c} q dq  J_0(qR) G^0_{AA} (q, E)
\label{gAA}
\end{equation}
and a momentum cutoff $q_c$ has been introduced. We get a similar expression for
$G^0_{AA} (0,R, E)$ and plugging these in Eq. \ref{chi}, we immediately get
\begin{equation}
\chi_{AA} (0,R) = I_{AA} (R) \times (1 + \cos [(K-K^\prime) \cdot R])
\label{chi-AA}
\end{equation}
with the prefactor
\begin{equation}
 I_{AA} (R) = - \frac{4}{\pi} \int^{E_F}  dE \  {\rm Im}  \ [g_{AA} (R, E)]^2.
\label{IAA}
\end{equation}

The large distance behavior of $J $  is controlled by small momentum states and one may try to evaluate this by taking the cutoff $q_c \rightarrow \infty$ for the ease of performing the integrals. In this case, the integral in Eq. \ref{gAA} can be expressed in terms of the modified Bessel's function of the second kind, viz.,
$g_{AA} = - 2 \pi \Omega_{BZ}^{-1} \times E /v_F^2 \times K_0 (i E R/v_F)$.\cite{Wang,Bena}
This in turn may be expressed in terms of Bessel and Neumann functions\cite{ASHandbook} with real arguments to yield the kernel
$
{\rm Im}  \ [g_{AA} (R, E)]^2 = (2 \pi^4) \Omega_{BZ}^{-2} v_F^{-2} R^{-2} y^2 J_0(y) Y_0 (y),
$
where $y = ERv_F^{-1}$. Thus Eq. \ref{IAA} becomes
\begin{equation}
I_{AA} (R) = \frac{8 \pi^3}{\Omega_{BZ}^2 v_F} R^{-3} \int_0^\infty dy y^2 J_0 (y) Y_0 (y).
\end{equation}

The $R^{-3}$ dependence clearly emerges; however, the integral does not converge. Following Saremi,\cite{Saremi} we multiply the integrand by a cutoff function $f(y, y_0)$, perform the integral, and take the limit $y_0 \rightarrow \infty$. We have tried three different cutoff functions, viz., $f(y, y_0) = \exp \ (-y/y_0), \ \exp \ (-y^2/y_0^2),$ or $y_0^2 (y_0^2 +y^2)^{-1}$, and in each case find the same limit: $\lim _{y_0 \rightarrow \infty}  \int_0^\infty dy y^2 J_0 (y) Y_0 (y) f(y, y_0) = 1/16$. Thus we immediately get
$I_{AA}(R) = 9 (64 \pi t )^{-1} a^3/R^3$, which leads to the exchange interaction for the same sublattice
\begin{equation}
J_{AA} = -C \times  \frac{1 + \cos ((K-K^\prime) \cdot R)}{(R/a)^3},
\label{Saremi}
\end{equation}
where $C \equiv - 9 \lambda^2 \hbar^2/ (256 \pi t)$ is a positive quantity, since  $t < 0 $ in graphene.
We note that the oscillatory factor is identical to the expression derived by Saremi;\cite{Saremi} however, we have an additional scaling factor of 3/2 for the magnitude of $J_{AA}$. From finite-size calculations, Black-Schaffer\cite{Annica} extracted a scaling factor different from ours or that of Saremi; however, we note that such factors are difficult to extract from numerical results, especially from finite-size calculations.

The expression for $J_{AA}$, Eq. \ref{Saremi}, is valid for all directions including the zigzag and the armchair directions and $K$ and $K'$ are any two adjacent Dirac points in the Brillouin zone. It is easy to see that while the oscillatory factor $1 + \cos ((K-K^\prime) \cdot R)$ repeats in triplets as 2, 1/2, 1/2, ... with distance $R$ along the zigzag direction, it is always two for the armchair direction. Because of this, the magnitude of $J_{AA}$ oscillates for the zigzag direction but not for the armchair direction, always however remaining ferromagnetic as required by the particle-hole symmetry.

\begin{figure}
\centering
\includegraphics[width=8.0cm]{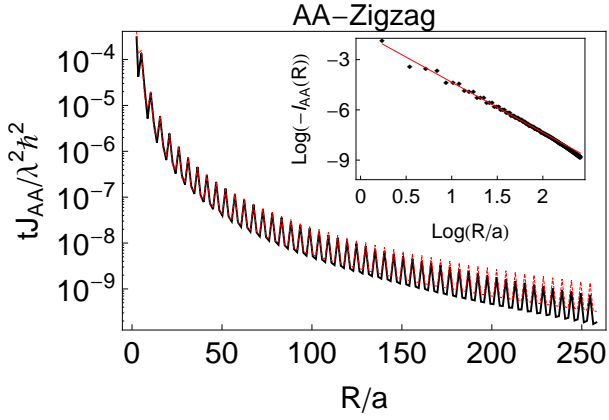}
\caption{(Color online) Exchange interaction $J_{AA}$ between two impurities on the same sublattice. Black lines are the results with the full tight-binding band structure, while the red lines indicate the long-distance behavior  as obtained from Eq. \ref{Saremi} using linear dispersion. The inset shows the log plot showing the long-distance $R^{-3}$ behavior, while there are noticeable differences for small $R$, especially visible in the inset. Note that since $t$ is negative for graphene, $J_{AA}$ is also negative indicating a ferromagnetic interaction.
}
\label{JAAzigzag}
\end{figure}

\begin{figure}
\centering
\includegraphics[width=8.0cm]{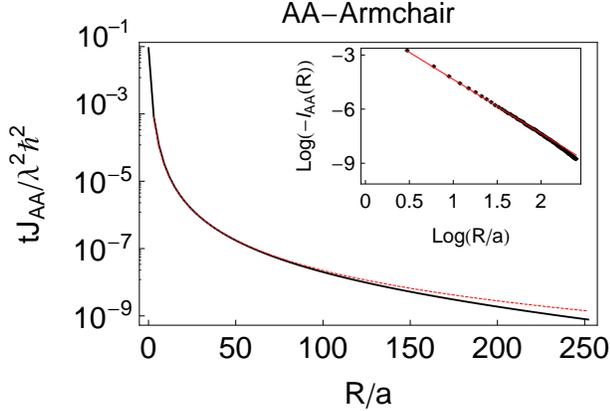}
\caption{(Color online) Same as Fig. \ref{JAAzigzag} for the armchair direction. For this direction, consistent with Eq. \ref{Saremi}, there are no oscillations in $J$ unlike the zigzag direction (Fig. \ref{JAAzigzag}).
}
\label{JAAarmchair}
\end{figure}

The calculated results for $J_{AA}$ using the full band structure following methods of Secs. II and III are shown in Fig. \ref{JAAzigzag} for the zigzag direction and in Fig. \ref{JAAarmchair} for the armchair direction. Both follow the long-distance behavior of Eq. \ref{Saremi} quite well beginning with surprisingly small distances.

\begin{figure}
\centering
\includegraphics[width=8.0cm]{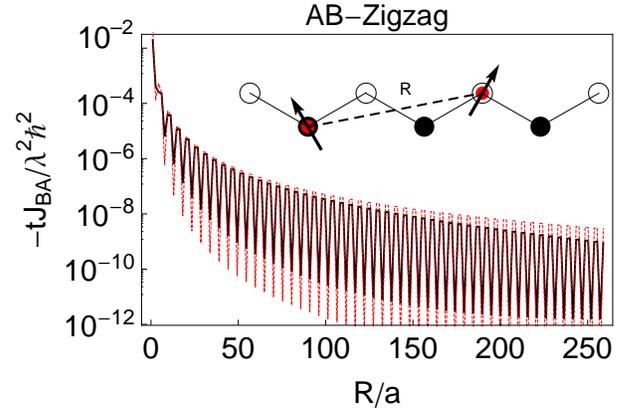}
\caption{(Color online) $J_{BA}$ for the zigzag direction. Black solid line is the result of the full calculation, while the red dashed line is  obtained from Eq. \ref{Saremi2}.
}
\label{Fig-AB-zigzag}
\end{figure}
%
\begin{figure}
\centering
\includegraphics[width=8.0cm]{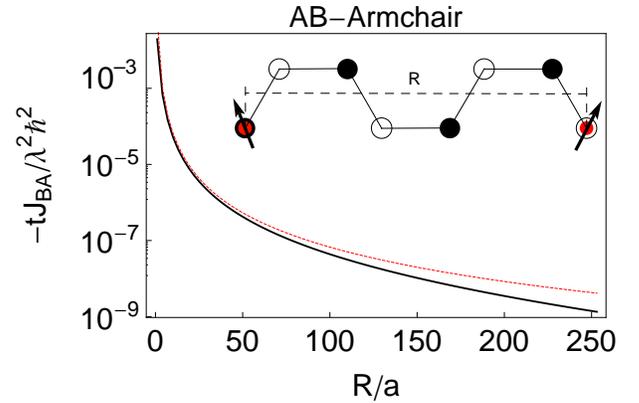}
\caption{(Color online) Same as Fig. \ref{Fig-AB-zigzag} for the armchair direction.}
\label{Fig-AB-armchair}
\end{figure}

\subsection{Magnetic impurities on two different sublattices}

We now turn to $J_{AB}$, where the two impurities are  located on different sublattices.  The  needed Green's functions, in the small $q$ limit, are obtained from Eq. \ref{GF-matrix}  to yield
\begin{eqnarray}
G^0_{BA}  (q+K_D, E)
 =  \pm \frac{v_F q \ e^{-i (\pi/3 \pm \theta_q)}}{(E+i \eta )^2 - v_F^2 q^2}
\end{eqnarray}
where $\pm$ signs are for the two Dirac points $K_D = K$ and $K^\prime$, respectively. Here, we have chosen a Brillouin zone that includes the $K_1$ and $K_2$ points and the corresponding phase factors in the Hamiltonian have been retained (see Fig. \ref{Fig-H}). The next step is to obtain $G^0_{BA} (R,0, E)$ by the momentum space integration using Eq. \ref{G-RE}. The same Jacobi-Anger expansion for $e^{i q\cdot R}$  (Eq. \ref{Jacobi}) may be used  as before except that now the extra phase factor $ e^{\pm i \theta_q }$ appears in the angle integral while performing the integration in Eq. \ref{G-RE}. Using the result
\begin{eqnarray}
\int_0^{2 \pi}d\theta_q     e^{\pm i \theta_q } \cos [n (\theta_q -\theta_R)] =\begin{cases}
0& \text{if $n\neq 1$},\\
 \pi e^{\pm i \theta_R}& \text{if $n=1$},
\end{cases}
\end{eqnarray}
where $\theta_R$ is the polar angle of the distance vector $R$ as defined in Fig. \ref{H},
we  get after some algebra, the result:
$
G^0_{BA} (R,0, E) = \alpha \
 g_{BA} (R, E),
$ where
$ \alpha =
e^{-i \pi/3} ( e^{i (K \cdot R - \theta_R)}-e^{i (K^\prime \cdot R + \theta_R)})  $
and
\begin{equation}
g_{BA} (R,E) = \frac {2 \pi i v_F}{\Omega_{BZ}} \int_0^{q_c} dq \frac {q^2  J_1(qR)}{(E+i \eta )^2 - v_F^2 q^2}.
\label{gBA}
\end{equation}
Similarly we find $G^0_{AB} (0, R, E)  = - \alpha^*  \
 g_{BA} (R, E)$.
 Finally, using Eq. \ref{chi}, the susceptibility becomes
\begin{equation}
\chi_{BA} (R,0) = I_{BA} (R) \times (1 + \cos [(K-K^\prime) \cdot R +\pi - 2\theta_R] )
\label{chi-BA}
\end{equation}
with the prefactor
\begin{equation}
 I_{BA} (R) = \frac{4}{\pi} \int^{E_F}  dE \  {\rm Im} \ [g_{BA} (R, E)]^2.
\label{IAB}
\end{equation}

We can now proceed to evaluate $J_{BA}$ in the long-distance limit in a fashion similar to the previous subsection. We find that
$g_{BA} =  2 \pi \Omega_{BZ}^{-1} \times E /v_F^2 \times K_1 (i E R/v_F)$, where $K_1$ is the first-order modified Bessel function of the second kind.
Expressing this in terms of Bessel and Neumann functions and using a cutoff function as before, we finally get
\begin{equation}
I_{BA} (R) = \frac{8 \pi^3}{\Omega_{BZ}^2 v_F} R^{-3} \times \lim_{y_0 \rightarrow \infty}   \int_0^\infty dy y^2 J_1 (y) Y_1 (y) f(y,y_0).
\end{equation}
The result for this integral is -3/16, so that  collecting all terms,  the exchange interaction becomes
\begin{equation}
J_{BA} = 3C \times
\frac {1 + \cos ((K-K^\prime) \cdot R +\pi - 2\theta_R) }    {(R/a)^3}.
\label{Saremi2}
\end{equation}

Note that the equation is valid for any direction of $R$ and for any choice of the two Dirac points $K$ and $K'$ (they have to be adjacent to each other of course, so that they are within a single unit cell in the reciprocal lattice), so long as we define the angle $\theta_R$ to be with respect to the chosen $K'-K$ vector. One can check that these equations yield the same results for $J$ for different equivalent directions $R$ as expected from symmetry.
	
	Note that $J_{BA}$ has the extra factor $\pi- 2 \theta_R$ in the argument of the cosine as compared to $J_{AA}$, which comes from the interference of the contributions to the Green's functions Eq. \ref{G-RE} from the two Dirac points. For the zigzag direction along $K'-K$,  the oscillatory part of Eq. \ref{Saremi2} agrees with the Black-Schaffer result,\cite{Annica} since the angle $\theta_R$ vanishes for large $R$. However, our result is valid for all directions and furthermore for the armchair directions, the angle $\theta_R$ is  never zero (see Fig. \ref{H}), so that this phase factor must be retained in Eq. \ref{Saremi2}.

	The oscillatory behaviors seen in the full calculations for $J_{AB}$ presented in Figs. \ref{Fig-AB-zigzag} and \ref{Fig-AB-armchair} are contained in Eq. \ref{Saremi2}. For the zigzag direction, taking $K-K' = 4 \pi (3 \sqrt 3 a)^{-1} (-1, 0)$ and $R = \vec {d_1} + \sqrt 3 a n \hat x$, where $n$ is an integer,  the oscillatory factor in Eq. \ref{chi-BA} becomes  $1+ \cos [(4n-1) \pi /3+2 \theta_R]$. In the limit $R \rightarrow \infty$, $\theta_R \rightarrow 0$, so that this factor repeats in the sequence of the triplet numbers: 0, 3/2, 3/2 as $R$ is increased. For a finite $R$, $\theta_R \ne 0$, so that we never get exactly the zero in the triplet, but rather a small number, which is faithfully reproduced in the full calculations shown in Fig. \ref{Fig-AB-zigzag}, where two values of $J$ are close in magnitude, while the next one is lower by about  three orders of magnitude. For the armchair direction, $R = 2^{-1} a (3n+1) (\sqrt 3, 1)$, $n$ being an integer and $\theta_R = \pi/6$, so that the oscillatory factor in Eq. \ref{chi-BA} is always two. The exchange interaction thus changes smoothly with distance without any oscillations as seen from Fig. \ref{Fig-AB-armchair}.

\subsection{Interaction between impurities on plaquettes}

\begin{figure}
\centering
\includegraphics[width=8.0cm]{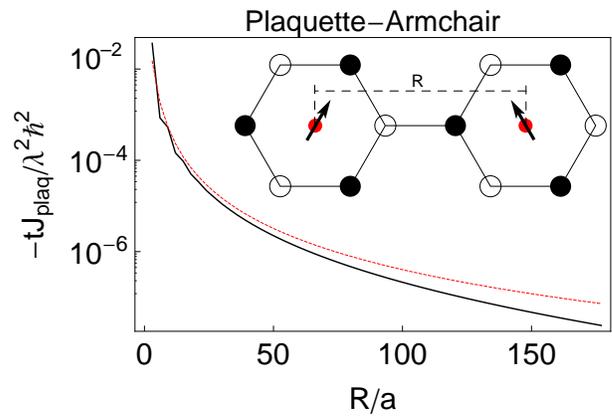}
\caption{Magnetic interaction $J$ for the hexagonal plaquette along the armchair direction.
}
\label{jplaqarm}
\end{figure}

\begin{figure}
\centering
\includegraphics[width=8.0cm]{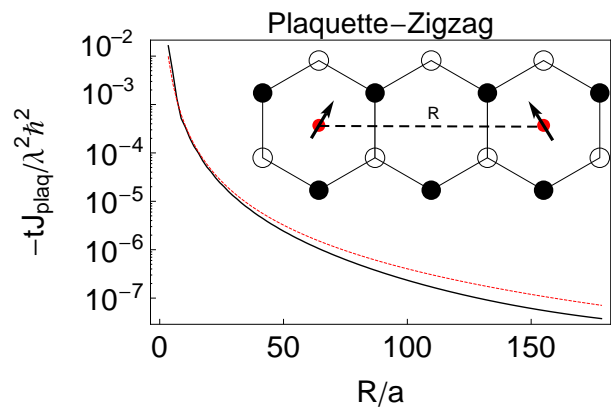}
\caption{Magnetic interaction $J$ for the hexagonal plaquette along the zigzag direction.
}
\label{jplaqzig}
\end{figure}

	The plaquette impurities, where the magnetic impurities are located at high-symmetry points rather than at single lattice sites,  are of interest because a number of atoms and small molecules may be favored to occupy such positions.
	For instance, the impurity may be located midway at a bond center and may interact with the two neighboring sites on the bond.
In these situations, the interaction term in Eq. \ref{hamil} may be replaced by
\begin{equation}
{\cal H}_{int} = - \lambda  \vec S_1 \cdot \sum_{p} \vec s_{p} - \lambda  \vec S_2 \cdot \sum_{p'} \vec s_{p'},
\end{equation}
where the summations are performed over the lattice sites  with which the impurity spins $\vec S_j$ interact (six sites if the impurity is located at the hexagon center and two, if it is on the bond center). Assuming that the interaction strength is small ($\lambda << |t|$), the interaction $J$ for these plaquette impurities may be obtained by simply summing over the individual site-site interactions already obtained in Sec. III, so that $J = \sum_{p, p'} J_{p p'}$.

\begin{figure}
\centering
\includegraphics[width=8.5cm]{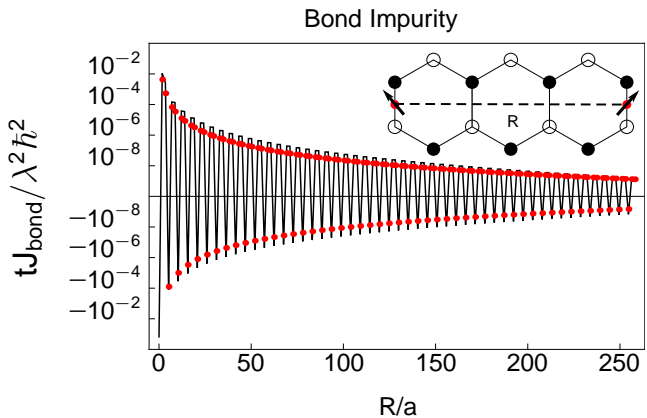}
\caption{ (Color online) Magnetic interaction $J_{bond}$ between two impurities located on bond centers obtained from the full calculation (solid line) and the linear-band long-distance limit Eq. \ref{Eq-jbond} (red dots).}
\label{jbond}
\end{figure}

These results are shown in Figs. \ref{jplaqarm} and \ref {jplaqzig} for the hexagonal plaquettes and in Fig. \ref{jbond} for impurities located on the bond centers. For the hexagonal plaquettes, as has been pointed out earlier,\cite{Saremi, Annica} the oscillating cosine factors present in the site interactions, $J_{AA}$ and $J_{AB}$, cancel out  both for the armchair and the zigzag hexagonal plaquette, leading to the single result valid for both cases:
\begin{equation}
J_{plaq} = 36 C \times  (R/a)^{-3},
\end{equation}
which is a net antiferromagnetic interaction.

For impurities located midway between the bond centers along the zigzag direction, we find an oscillating interaction, in the long-distance limit,
\begin{equation}
J_{bond} (R) = 4 C    \times
\frac {1 - 2 \cos ((K-K^\prime) \cdot R ) }    {(R/a)^3}.
\label{Eq-jbond}
\end{equation}
As distance $R$ is increased, the numerator changes with the repeat sequence of -1, 2, and 2, leading to an interaction that is antiferromagnetic every third site and ferromagnetic otherwise. The results from the full calculation is compared to the long-distance limit result, Eq. \ref{Eq-jbond}, in Fig. \ref{jbond}.

\section{Summary and Discussions}

In summary, we studied the RKKY interaction between the magnetic impurities in the honeycomb lattice by evaluating the Green's functions for the tight-binding Hamiltonian by the direct summation method, which worked well for all distances but is  computationally slow, or the Horiguchi recursive technique, which is a fast method, but has stability problem for large $R$. For distances, where both methods worked, the results agreed with each other quite well. These methods are complementary to the finite-lattice calculations\cite{Annica}, however, the direct summation method allows the calculation of $J$ for much larger distances with modest computational efforts. By carefully considering the phase factors of the wave functions around the Dirac cones, we have also obtained the analytical long-distance limits of $J$, Eqs. \ref{Saremi} and \ref{Saremi2}, which are, although similar in form to previous results,\cite{Saremi, Annica} have important corrections in terms of additional phase factors in the oscillating term. All such phase factors were faithfully reproduced in our numerical calculations of $J$ using the full tight-binding band structure. We found that the long-distance  limit is reached for quite small distances, of the order of a few lattice constants.

In addition to the nearest-neighbor model, we have also studied the effect of the further-neighbor electron hopping, but these produced negligible differences as might be expected, since the strengths of the higher-neighbor hoppings are quite small in graphene.\cite{Nanda-Graphene} For the hexagonal plaquette impurities, $J$ is always antiferromagnetic, while for the bond impurities, the sign oscillates. Given that the magnitude of $J$ falls off quite rapidly with distance, nearest-neighbor $J$ would dominate, so that spin chains based on hexagonal plaquette sites are prediced to be antiferromagnetic, while those based on the bond sites would be ferromagnetic.

In trying to design an experimental system to observe the RKKY interaction, one must carefully select a proper system.  At first sight, it might appear that a magnetic adatom such as Co or Fe deposited on top of the graphene sheet would interact via the RKKY interaction. However, in addition to introducing the needed localized magnetic moments from the $d$ electrons, the outermost $s$ electrons of the adatom are transferred to the host, where they are added to either the conduction band or they may form weakly localized states around the adatom site. These extra electrons will modify the RKKY interaction. The challenge is therefore to come up with a system, perhaps a simple molecule, that has a magnetic moment which interacts with the graphene lattice, but one that does not alter the electronic structure by contributing extra electrons to the graphene sheet.  On the other hand, scaling arguments\cite{Fradkin} as well as renormalization group calculations\cite{Castro} indicate the lack of a Kondo effect below the critical coupling $J_c \approx 3.5$  eV, which is quite strong, so that the RKKY interaction should dominate.

This work was supported by the U. S. Department of Energy through Grant No.
DE-FG02-00ER45818. MS thanks Ali Tayefeh Rezakhani for useful discussions.


\end{document}